# Two Information Entropies of Square-Lattice ±J Ising Models

Xiao Xu (徐晓), Zhongquan Mao (毛忠泉), Xi Chen (陈熹)*

Department of Physics, South China University of Technology, Guangzhou 510640, China

**Abstract:** Two information entropies, $S_{IB}$ based on the local bonding energy configurations and $S_{IS}$ based on the local spin configurations, are applied in the square-lattice ±J Ising systems with Monte Carlo simulations. With $S_{IB}$ and $S_{IS}$, the spin glass states can be distinguished from paramagnetic states clearly. The results reveal that the nature of spin glass states is of ordering in bonding energy and disordering in spin orientations. The consistence of $S_{IB}$ and the thermodynamic entropy is found.

* xichen@scut.edu.cn

Spin glass (SG), though has been studied for decades as a prototype of the systems with frustration and disorder[1,2,3], still puzzles us: its spin configuration keeps disordered at low temperature while the diminishment of its entropy occurs [1], indicating SG is an ordered state. According to Landau's theory, the correlation length [4,5] is a unified order parameter for an order-to-disorder phase transition. In SG, however, the correlation length is hard to calculate due to the disordered spin configuration [6]. Therefore Edwards and Anderson had introduced an order parameter $q_{EA}$ for SG. [7,8,9] However this order parameter vanishes in some systems when the system size and the observation time trend to infinite. [3,10,11] Experimentally, the cusp of the AC magnetic susceptibility [12,13], which could be deduced from $q_{EA}$ theoretically [7], and the maximum of heat capacity [14] are used to identify the critical temperature $T_c$ of the SG transition. However, the peak position of the AC magnetic susceptibility shifts to higher $T_c$ with the frequency increasing [15]; and the maximum of heat capacity [7], derived from Ehrenfest's phase classifications, does not correspond to $T_c$ for the problem of so-called 'configurational entropy' in SG [16]. The determination of the spin glass state is still a challenge.

It is well recognized that entropy is the measure of disorder in physics. Hence the thermodynamic entropy could be used to describe the SG disorder. In practice, it is difficult to estimate the Boltzmann-Gibbs (BG) entropy of a large thermodynamic system because of the tremendous number of microstates. [17,18] Fortunately, a kind of *information entropy* can be calculated based on the probability distribution of a certain characteristics for a system. [19,20] It is found that, in ferromagnetic systems, the information entropies show peaks near $T_c$. [21,22,23] Moreover, several information entropies have been used to study the order-to-disorder transitions. [24,25] These hint strongly that information entropies corresponding to the features of SG may provide a new sight to the puzzle of disordered spin configurations in ordered state.

According to the information theory, the information entropy can be defined mathematically based on one discrete probability distribution of some feature and represents the disorder of this feature no matter what the feature is. [19,20] In this letter, we defined two information entropies, $S_{IB}$, based on the local bonding energy configuration, and $S_{IS}$, on the local spin configuration, respectively in the square-lattice $\pm J$ Ising systems [1,26,27,28] under zero-field conditions and compared them with the thermodynamic entropy $S_T$. By extensive Monte Carlo (MC) simulations, we found that both ferromagnetic (FM) and antiferromagnetic (AF) states can be distinguished from paramagnetic (PM) phase by either $S_{IB}$ or $S_{IS}$, but only $S_{IB}$ can be used to differentiate SG state from PM phase. Furthermore, we demonstrate that $S_{IB}$ is almost linear with $S_T$. The consistence of $S_{IB}$ and $S_T$ suggests that the order of SG is from the ordered bonding energy configurations.

Consisting of a central spin, four nearest spins and four bonds between them, a basic cell (BC) is chosen to estimate the information entropies for a square-lattice $\pm J$ Ising system with $N \times N$ Ising spins from local zones. According to the spin configurations, there are $2^5 = 32$ states with 'up' or 'down' directions for each spin, while according to the bonding energy configuration, there are $2^4 = 16$ states with 'high' or 'low' energy state for each bond. Under the periodic boundary condition, $N \times N$ BCs can be extracted.

For a local spin configuration state $i$, the probability of finding this state, $p_{S,i}$, is

$$p_{S,i} = \frac{N_{S,i}}{N_{\Sigma}} \tag{1}$$

where $N_{S,i}$ is the number of BCs on State $i$ of spin configurations extracted from the system and $N_{\Sigma}$ is the total number of BCs, i.e. $N_{\Sigma} = N \times N$. Then the information entropy of the system on spin configurations is,

$$S_{IS} = -\sum_{i=1}^{n_s} p_{S,i} \ln(p_{S,i}) \tag{2}$$

where $n_s (=32)$ is the total number of states of the spin configurations. Similarly, the information entropy on bonding energy configurations can be written as

$$S_{IB} = -\sum_{i=1}^{n_B} p_{B,i} \ln(p_{B,i}) \tag{3}$$

where $p_{B,i} = \frac{N_{B,i}}{N_{\Sigma}}$ is the probability of finding State $i$ of bonding energy configurations, $N_{B,i}$ is number of BCs staying on State $i$ of bonding energy configurations in a system, and $n_B(=16)$ is the total number of states of the bonding energy configurations. From the information theory perspective, $S_{IS}$ represents the disorder of spin configuration of the system since it is calculated from the local spin configurations, and $S_{IB}$ represents the disorder of bonding energy.

On the other hand, according to the thermodynamic cardinal equation, under zero-field condition, with the change in the system's volume ignored, the derivative of thermal entropy $S_T$ per spin to temperature $T$ is

$$\frac{1}{N_\Sigma}\frac{dS_T}{dT} = C/T \tag{4}$$

where C is the heat capacity per spin. Obviously, one can compare $C/T$ with $\frac{dS_{IS}}{dT}$ or $\frac{dS_{IB}}{dT}$ directly to find the relationship between $S_T$ and $S_{IS}$ or $S_{IB}$.

The Monte Carlo simulations were performed on the square-lattice Ising systems with the nearest-neighbor coupling $J_{ij}$. Under zero-field condition, the Hamiltonian of the system is,

$$H = -\sum_{<ij>} J_{ij}\, \sigma_i\, \sigma_j \tag{5}$$

where $\sigma_i$ is the Ising spin, and the sums $<ij>$ runs over all the nearest-neighbor pairs of spins. The interaction energy $J_{ij}$ takes $\pm J$ ($J > 0$) randomly with the probability distribution,

$$P(J_{ij}) = r\delta(J_{ij} + J) + (1-r)\delta(J_{ij} - J) \tag{6}$$

where r is the ratio of the number of AF bonds to that of all the bonds in a system. Here $r = 0$ for a typical FM system, $r = 1$ for a typical AF system and $r = 0.5$ for a standard EA glass model. Then the heat capacity (per spin) is,

$$C = \frac{1}{N_\Sigma}\frac{dH}{dT} \tag{7}$$

All Monte Carlo simulations are performed with a sequential heat-bath algorithm [29,30]. As the spin glass is an non-equilibrium state, the system should always evolve [10], which should affect the stability of $S_{IB}$ and $S_{IS}$. Hence we check the evolutions of $S_{IB}$ and $S_{IS}$ with $r = 0.5$ at two temperature points near $T_c$ with $1.6\times10^6$ MC steps. No evolving phenomenon for $S_{IB}$ and $S_{IS}$ could be observed after 3000 MC steps. Thus $10^4$ MC steps are employed for thermal equilibrium at each temperature for all simulations. The temperature $T$ is reduced by $J/k_B$, where $k_B$ is the Boltzmann constant. To simplify the calculation, we set $J=1$.

The temperature $T$ and dilution ratio $r$ dependence of $S_{IS}$ and $S_{IB}$ are displayed in Fig 1 (a) and (b) respectively with $r = 0, 1/21, 2/21, \ldots, 1$. Here the system size is $100\times100$ spins and the result is averaged over 64 simulations with different initial spin configurations and bond distributions at a high temperature. In both (a) and (b), $S_{IB}$ and $S_{IS}$ decreases rapidly when a system transits into an AF or FM state from the PM states. Only $S_{IB}$ can give the SG states a lower terrace and the PM states a higher one while $S_{IS}$ shows an undivided platform for SG and PM states. Furthermore, $S_{IB}$ is symmetric along $r = 0.5$, while the symmetry of $S_{IS}$ is broken since the two orientations of spins in AF systems appear equally likely but not in FM systems.

As expected, both $S_{IS}$ and $S_{IB}$ are relatively large in PM states since spins are independently disordered of each other. In FM and AF states, the spins are strongly correlated and well aligned. Hence both $S_{IS}$ and $S_{IB}$ are relatively small. Interestingly, in SG states, the values of $S_{IS}$ are close to those in PM states, while the values of $S_{IB}$ are smaller than those in PM states but clearly larger than those in FM and AF states. It is well known that the spins are correlated in SG despite of the disordered spin configurations. Our results clearly demonstrate the nature of energy order in the spin disorders in SG.

$\frac{dS_{IB}}{dT}$, $\frac{dS_{IS}}{dT}$ and $C/T$ are calculated to investigate the relationship between the thermodynamic entropy $S_T$ and the information entropies $S_{IB}$ and $S_{IS}$. As shown in Fig.2, $C/T$ is almost proportional to $\frac{dS_{IB}}{dT}$ while it is apparently different from $\frac{dS_{IS}}{dT}$, and hence $S_{IB}$ and $S_T$ behave similarly.

As shown in Ref. 26, the entropy of SG states approaches a non-zero constant when the temperature approaches zero, and at very high temperatures the entropy of PM approaches a higher constant. A dramatic change of entropy occurs between SG and PM states. It makes sense to use the maximum of C/T, i.e. $\frac{1}{N_\Sigma}\frac{dS_T}{dT}$ as shown in Eq. (4), for determining the phase transition temperature $T_c$.

The precise derivatives of different entropies to temperature for the FM system ($r = 0$) with 400×400 spins and the SG system ($r = 0.5$) with 200×200 spins are displayed in Fig. 3 (a) and (b) respectively. To accelerate the simulations, the sequential Metropolis algorithm [31] instead of the heat-bath algorithm is used to update the spins in the FM system. Here the averages are preformed over 90 simulations for the FM system and 60 simulations for SG system.

For the square-lattice Ising FM magnet ($r = 0$), the analytical solution gives $T_c = 2.2692$. [32,33] By fitting the magnetic moment data with $M = M_0\left(1 - T/T_C\right)^{1/8}$, we obtain $T_c = 2.265 \pm 0.004$, very close to the analytical result. The peaks of $\frac{dS_{IS}}{dT}$, $\frac{dS_{IB}}{dT}$ and C/T are at 2.267 ± 0.005, 2.268 ± 0.005 and 2.269 ± 0.005 as shown in Fig.3(a). Obviously all these peak positions can not be distinguished from the analytical result in the context of experimental precise. Not surprisingly, the $\frac{dS_{IB}}{dT}$ is almost proportional to the C/T. It should be noticed that the $\frac{dS_{IS}}{dT}$ is sharper than the $\frac{dS_{IB}}{dT}$, since the strong ergodic breaking gives rise to the symmetry breaking of spin orientations in ferromagnetic systems [34].

For the typical EA SG ($r = 0.5$) shown in Fig.3 (b), the $C/T$ and $\frac{dS_{IB}}{dT}$ behave similar, but their peak positions are not coincident with each other: 1.11±0.03 for $C/T$ and 1.24±0.03 for $\frac{dS_{IB}}{dT}$ which are close to the freezing temperature $T_f \approx 1.25$ by EA parameter $q_{EA}$ [26] and the critical temperature $T_c \approx 1.3$ with the replica number equals to 2 [35]. When we enlarge the basic cell to 3×3 spins with 12 bonds connecting neighbor spins, the $\frac{dS_{IB}}{dT}$ vs. $T$ curve shifts more close to $C/T$ curve (not shown in this paper) and the phase transition point $T_c \approx 1.14$. It means by increasing the size of a basic cell, one can reduce the disparity between the information entropy and the thermodynamic entropy. A big cell, however, will bring up a tremendous expansion of the calculation.

Why does $S_{IB}$, but not $S_{IS}$, behave similar to the thermal entropy $S_T$? The reason is that both $S_T$ and $S_{IB}$ relate to the energy distribution of the system while $S_{IS}$ only relates to the spin configurations. When the spin configurations strongly rely on the energy distribution such as in the FM or AF phase, $S_{IS}$ has similar behavior as $S_T$. However in SG state, due to the frustrations, the corresponding relationship between spin configurations and energy distributions misses. Hence $S_{IS}$ can not reveal the changes of $S_T$.

Here we interpret the linear relationship between $S_T$ and $S_{IS}$ as following.

The Hamiltonian of a system can be carried out with,

$$H = \frac{1}{2} N_\Sigma \cdot \sum_{i=1}^{n_B} p_{B,i} H_i \tag{8}$$

where $H_i$ is the Hamiltonian of a BC in State $i$ of bonding energy configurations. For the convenience of calculation, set '0' for the energy of the bond staying on the low level and '2' for that on the high level. Then $H_i$ equals to 2 times of the number of high level bonds exist in a BC of State $i$.

From Eq. (4), (7) and (8), one has,

$$\frac{dS_T}{dT} = N_\Sigma \cdot \sum_{i=1}^{n_B} \frac{H_i}{2T} \frac{dp_{B,i}}{dT} \qquad (9)$$

Considering the information entropy on bonds defined as Eq. (3), one has,

$$\frac{dS_{IB}}{dT} = -\frac{d\left(\sum_{i=1}^{n_B} p_{B,i}\right)}{dT} - \sum_{i=1}^{n_B} \frac{dp_{B,i}}{dT} \ln p_{B,i}$$
$$= \sum_{i=1}^{n_B} \frac{dp_{B,i}}{dT} \ln(1/p_{B,i}) \qquad (10)$$

If the $\ln(1/p_{B,i})$ is linear to $\frac{H_i}{2T}$, i.e.

$$\ln(1/p_{B,i}) = a\frac{H_i}{2T} + b(T) \qquad (11)$$

where $a$ is the scaling factor and $b$ is the intercept, one has

$$\frac{dS_{IB}}{dT} = a\sum_{i=1}^{n_B} \frac{dp_{B,i}}{dT} \frac{H_i}{2T} + b(T)\sum_{i=1}^{n_B} (\frac{dp_{B,i}}{dT})$$
$$= a\sum_{i=1}^{n_B} \frac{dp_{B,i}}{dT} \frac{H_i}{2T} + b(T)\frac{d\left(\sum_{i=1}^{n_B} p_{B,i}\right)}{dT}$$
$$= a\sum_{i=1}^{n_B} \frac{dp_{B,i}}{dT} \frac{H_i}{2T} \qquad (12)$$

Camparing Eq. (12) with Eq. (9), we have

$$\frac{dS_T}{dT} \propto \frac{dS_{IB}}{dT} \qquad (13)$$

If the BCs are independent of each other, the energy of BCs follows Boltzmann distribution, and hence Eq. (13) holds. However this relationship could be changed since the couplings between the BCs can affect the distributions of bonds. Interestingly our simulations demonstrate $\frac{dS_T}{dT}$ is almost proportional to $\frac{dS_{IB}}{dT}$. This implies that the couplings between BCs have no remarkable influence on the slopes $a$ given by Eq. (11) near phase transition point where the information entropy varies dramatically. Shown as an example in Fig.4, for $r = 0.5$, the slopes $a$ of fitting lines for the functions $\ln(1/p_{B,i})$ of $H_i/2T$ are about 2.0 near the phase transition point, which means $\frac{dS_T}{dT}$ is proportional to $\frac{dS_{IB}}{dT}$ in the critical zone. The slope $a$ deviates from 2.0 at low temperature, e.g. $a \approx 0.9$ at $T = 0.1$. However $\frac{dS_T}{dT}$ and $\frac{dS_{IB}}{dT}$ approach zero at low temperature, which have weak affection on the linear relationship between $S_T$ and $S_{IB}$. We checked all systems for $r = 0$ to 1 and found that the slope $a$ maintains a constant near the critical temperature. Interestingly, $\ln(1/p_{B,i})$ has a better linearity with $H_i/2T$ above $T_c$ than that below $T_c$ although the fitting slope $a$ keeps changeless. It seems that a constant slope $a$ at the vicinity of $T_c$ is the key factor for the linearity between $S_{IB}$ and $S_T$.

Our derivation suggest that if one could find a local energy distribution $p_{B,i}$ with $\ln p_{B,i}$ approximately linear to $aH_i/T$ at the vicinity of $T_c$, the information entropy $S_{IB}$, which is linear to the thermodynamic entropy $S_T$, could be always calculated.

In conclusion, we define the information entropies $S_{IS}$ and $S_{IB}$ based on the local spin configurations and local bonding energy configurations respectively, and apply them to the $\pm J$ systems. We clarify that SG is a state disordered in spin configurations and ordered in bonding energy configurations. It is found that $S_{IB}$ is nearly linear to the thermodynamic entropy when $\ln p_{B,i}$ is approximately linear to $aH_i/T$ at the vicinity of $T_c$.

Thanks our colleagues, Prof. Jiang Zhang, Prof. Hongjun Quan, Prof. Yujun Zhao, and

Prof. Xiaobao Yang for the discussions. Fruitful discussions with Prof. Zexian Cao in Institute of Physics, Chinese Academy of Sciences, Beijing, are gratefully acknowledged. This work is supported by NSFC Grant Nos. 11174083 and 11304098.

Figure Captions:

Figure 1 The contours of (a) $S_{IS}$ and (b) $S_{IB}$ as the functions of $r$ and $T$. FM, SG, AF and PM states are marked in (b) based on different color zones, while SG and PM can not be distinguished from each other in (a).

Figure 2 The C/T (blue solid lines), $dS_{IS}/dT$ (green dash dot lines) and $dS_{IB}/dT$ (red dash lines) as the functions of $T$ under $r =$ 2/21 (a), 5/21 (b), 9/21 (c), and 19/21 (d).

Figure 3 (a) Normalized $C/T$, $\dfrac{dS_{IS}}{dT}$ and $\dfrac{dS_{IB}}{dT}$ for $r = 0$, and (b) $C/T$ and $\dfrac{dS_{IB}}{dT}$ for $r = 0.5$ as the function of temperatures. In (b), 10000 MC steps are performed for averages after 10000 MC steps for thermal equilibrium at each temperature.

Figure 4 The relationships between $\ln(1/p_{B,i})$ and $H_i/2T$ for $r = 0.5$ at the temperature range between 0.6 and 1.8. There are 1 state for $H_i = 0$, 4 states for $H_i = 2$, 6 states for $H_i = 4$, 4 states for $H_i = 6$, and 1 state for $H_i = 8$ at each temperature point, and for $H_i = 4$, the markers split into two branches, the lower branch for the states with the two high level bonds staying on the symmetric positions about the central spin, and the upper branch for the others. Four lines, labeled as A, B, C and D, are the linear fitting lines for $\ln(1/p_{B,i})$ with different $H_i$ at $T = 1.8(■)$, $1.4(♦)$, $1.0(●)$ and $0.6(▲)$ respectively. The slopes of the fitting lines are 2.0, 2.0, 2.0, and 1.8, respectively. Here all data for $p_{B,i} < e^{-10}$ are ignored due to a large estimation error in the 200×200 systems.

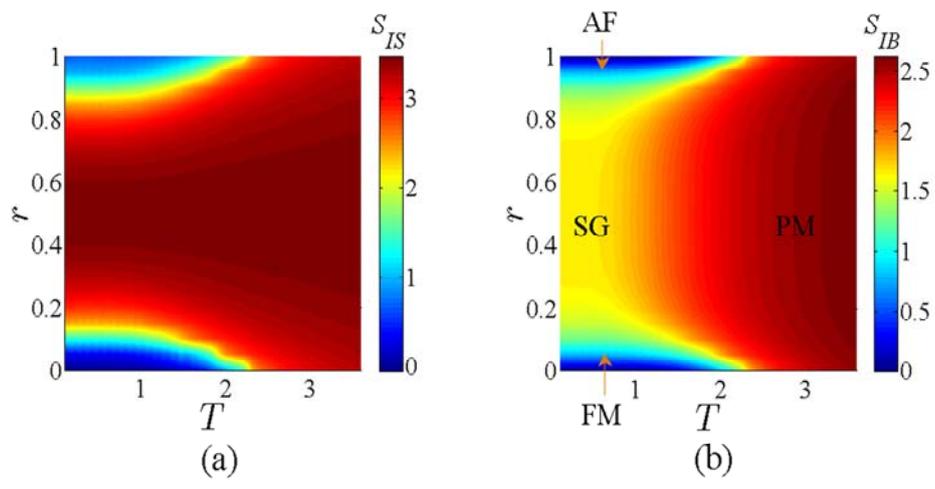

Fig. 1   Xiao Xu et al.

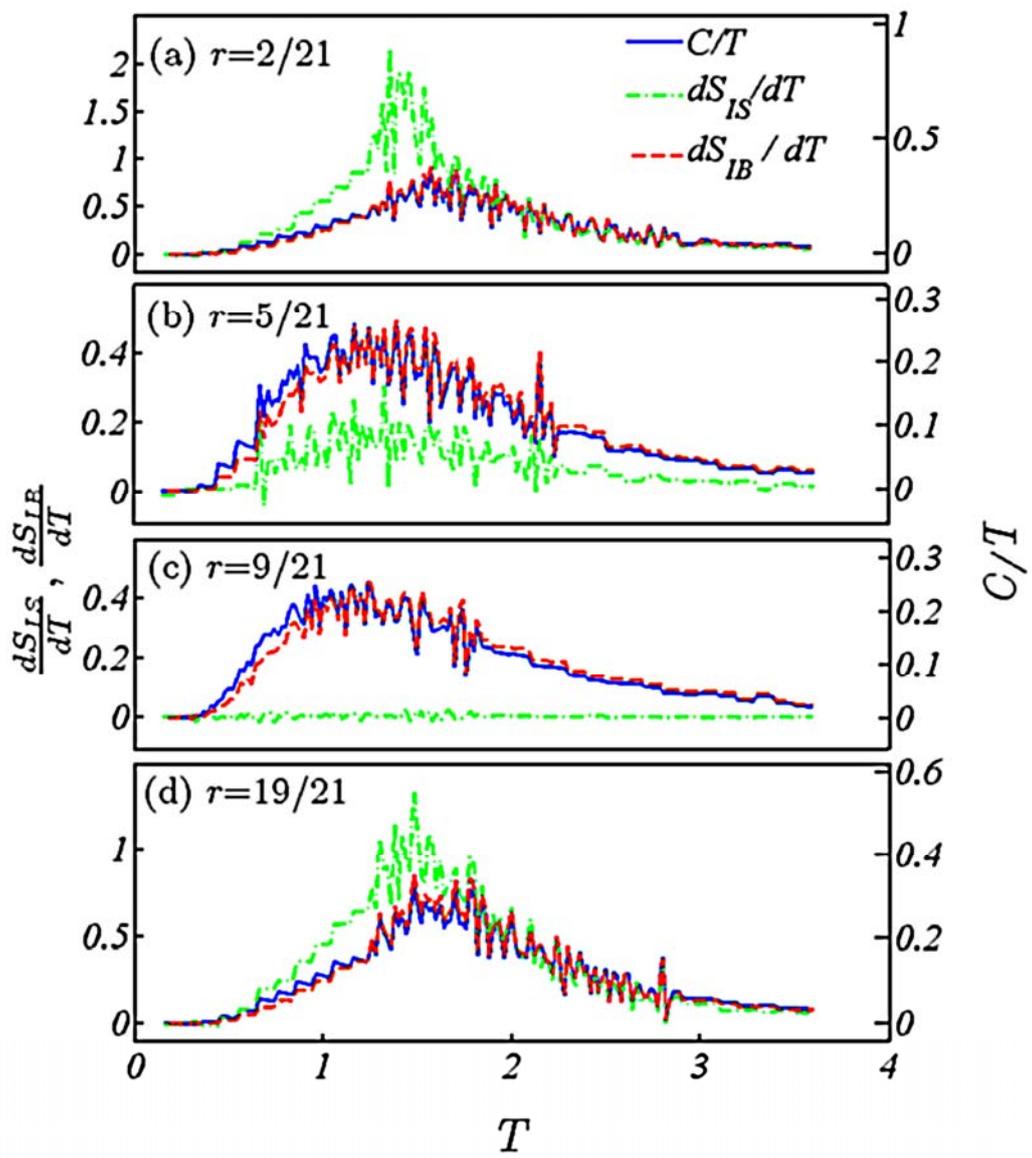

Fig. 2　Xiao Xu et al.

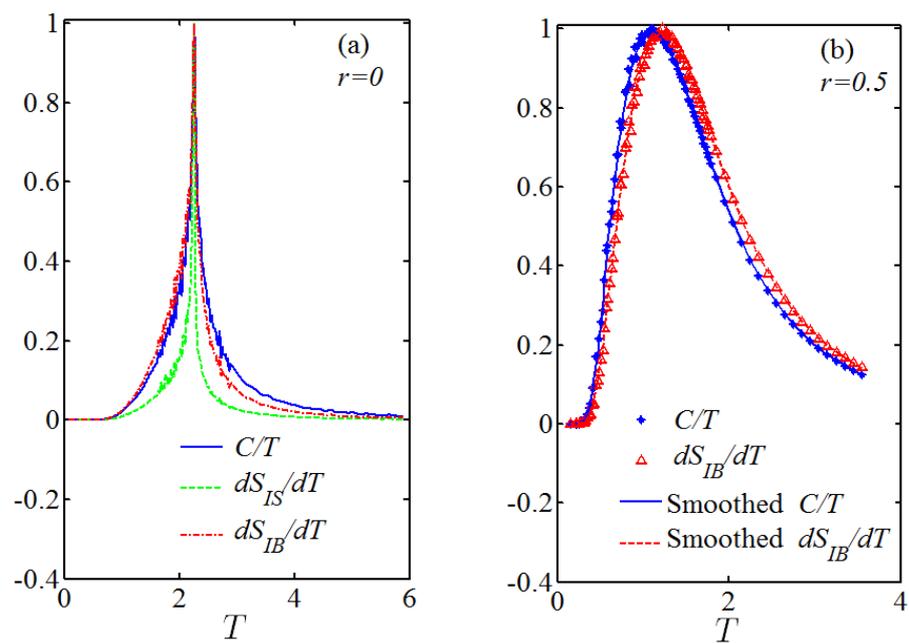

Fig.3　Xiao Xu et al.

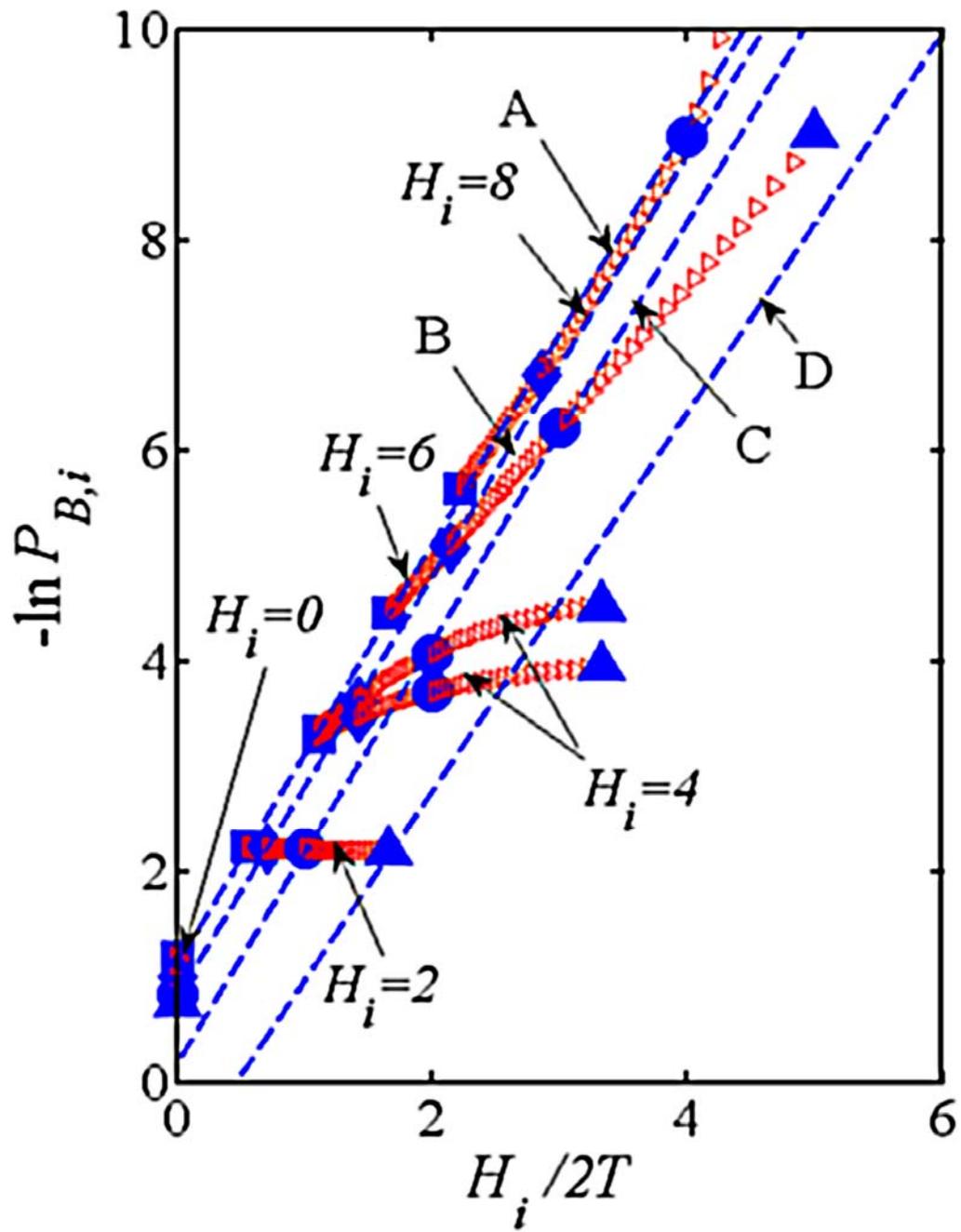

Fig.4    Xiao Xu et al.